\documentclass[twocolumn]{aastex7}
\usepackage{xspace}

\newcommand{\NEIDSM}{$\texttt{NEIDSpecMatch}$\xspace}

\newcommand{\PSUARC}{Astrobiology Research Center, 525 Davey Laboratory, 251 Pollock Road, Penn State, University Park, PA 16802, USA}

\newcommand{\PSUAA}{Department of Astronomy \& Astrophysics, 525 Davey Laboratory, The Pennsylvania State University, University Park, PA 16802, USA}
\newcommand{\PSUCEHW}{Center for Exoplanets and Habitable Worlds, 525 Davey Laboratory, The Pennsylvania State University, University Park, PA 16802, USA}
\newcommand{\UA}{Steward Observatory, University of Arizona, 933 N.\ Cherry Ave, Tucson, AZ 85721, USA}

\newcommand{\Macquarie}{School of Mathematical and Physical Sciences, Macquarie University, Balaclava Road, North Ryde, NSW 2109, Australia}

\newcommand{\JPL}{Jet Propulsion Laboratory, California Institute of Technology, 4800 Oak Grove Drive, Pasadena, CA 91109, USA}

\newcommand{\UCI}{Department of Physics \& Astronomy, The University of California, Irvine, Irvine, CA 92697, USA}
\newcommand{\Carleton}{Carleton College, One North College St., Northfield, MN 55057, USA}

\begin{document}

\title{\NEIDSM: stellar parameter estimation with NEID spectra using an empirical library}
\author[0000-0002-7127-7643]{Te Han} 
\affiliation{\UCI}
\email{teh2@uci.edu}

\author[0000-0003-0149-9678]{Paul Robertson}
\affil{\UCI}
\email{paul.robertson@uci.edu}

\author[0000-0003-4835-0619]{Caleb I. Ca\~nas}
\altaffiliation{NASA Postdoctoral Fellow}
\affiliation{NASA Goddard Space Flight Center, 8800 Greenbelt Road, Greenbelt, MD 20771, USA }
\email{c.canas@nasa.gov}

\author[0000-0001-7409-5688]{Gudmundur Stefansson}
\affil{Anton Pannekoek Institute for Astronomy, University of Amsterdam, Science Park 904, 1098 XH Amsterdam, The Netherlands}
\email{gummiks@gmail.com}

\author[0000-0001-8401-4300]{Shubham Kanodia}
\affil{Earth and Planets Laboratory, Carnegie Science, 5241 Broad Branch Road, NW, Washington, DC 20015, USA}
\email{skanodia@carnegiescience.edu}

\author[0000-0001-8720-5612]{Joe P.\ Ninan}
\affil{Department of Astronomy and Astrophysics, Tata Institute of Fundamental Research, Homi Bhabha Road, Colaba, Mumbai 400005, India}
\email{indiajoe@gmail.com}

\author[0000-0003-0353-9741]{Jaime A. Alvarado-Montes}
\affil{\Macquarie}
\affil{Astrophysics and Space Technologies Research Centre, Macquarie University, Balaclava Road, North Ryde, NSW 2109, Australia}
\email{jaime-andres.alvarado-montes@hdr.mq.edu.au}

\author[0000-0003-4384-7220]{Chad F.\ Bender}
\affil{\UA}
\email{cbender@arizona.edu}

\author[0000-0002-3610-6953]{Jiayin Dong}
\affiliation{Center for Computational Astrophysics, Flatiron Institute, 162 Fifth Avenue, New York, NY 10010, USA}
\affiliation{Department of Astronomy, University of Illinois at Urbana-Champaign, Urbana, IL 61801, USA}
\email{jdong@flatironinstitute.org}

\author[0000-0002-3853-7327]{Rachel Fernandes}
\altaffiliation{President’s Postdoctoral Fellow}
\affiliation{\PSUAA}
\affiliation{\PSUCEHW}
\email{rbf5378@psu.edu}

\author[0000-0002-5463-9980]{Arvind F.\ Gupta}
\affil{U.S. National Science Foundation National Optical-Infrared Astronomy Research Laboratory, 950 N.\ Cherry Ave., Tucson, AZ 85719, USA}
\email{arvind.gupta@noirlab.edu}

\author[0000-0003-1312-9391]{Samuel Halverson}
\affil{\JPL}
\email{samuel.halverson@jpl.nasa.gov}

\author[0000-0001-9626-0613]{Daniel M. Krolikowski}
\affil{\UA}
\email{krolikowski@arizona.edu}

\author[0000-0002-9082-6337]{Andrea S.J.\ Lin}
\affil{Department of Astronomy, California Institute of Technology, 1200 E California Blvd, Pasadena, CA 91125, USA}
\email{andrealin628@gmail.com}

\author[0000-0001-9596-7983]{Suvrath Mahadevan}
\affil{\PSUAA}
\affil{\PSUARC}
\affil{\PSUCEHW}
\email{suvrath@gmail.com}

\author[0000-0003-1324-0495]{Leonardo A. Paredes}
\affil{\UA}
\email{lparedes@arizona.edu}

\author[0000-0001-8127-5775]{Arpita Roy}
\affil{Astrophysics \& Space Institute, Schmidt Sciences, New York, NY 10011, USA}
\email{arpita308@gmail.com}

\author[0000-0002-4046-987X]{Christian Schwab}
\affil{\Macquarie}
\email{mail.chris.schwab@gmail.com}

\author[0000-0002-4788-8858]{Ryan C. Terrien}
\affil{\Carleton}
\email{rterrien@carleton.edu}

\begin{abstract}
We introduce \NEIDSM, a tool developed to extract stellar parameters from spectra obtained with the NEID spectrograph. \NEIDSM is based on \texttt{SpecMatch-Emp} and \texttt{HPFSpecMatch}, which estimate stellar parameters by comparing the observed spectrum to well-characterized library spectra. This approach has proven effective for M dwarfs. Utilizing a library of 78 stellar spectra covering effective temperatures from $3000-6000$\,K, \NEIDSM derives key parameters, including effective temperature, metallicity, surface gravity, and projected rotational velocity. Cross-validation shows median uncertainties of $\sigma_{T_{\mathrm{eff}}} = 115\,\mathrm{K}$, $\sigma_{[\mathrm{Fe/H}]} = 0.143$, and $\sigma_{\log g} = 0.073$ across 49 orders. We showcase its application by fitting the spectrum of an M-dwarf and discuss its utility across a wide range of spectra observed with NEID. \NEIDSM is \texttt{pip}-installable. 
\end{abstract}

\section{Introduction} \label{sec:intro}
Precise stellar parameter estimation is essential for characterization of stars and their planets. Methods have been developed for different spectral types but many struggle on late-K and M dwarfs. \cite{Yee} introduced the \texttt{SpecMatch-Emp} algorithm that directly compares observed spectra to a library of 404 FGKM stars with well-determined stellar parameters, first applied to spectra taken with HIRES on Keck. By taking $\chi^2$ values between each library spectrum and the target observation, the most similar library stars are selected. These library stars are then used to derive a composite spectrum and derive the observed star's parameters ($T_{\text{eff}}$, $R$, [Fe/H]). Although the method of \texttt{SpecMatch-Emp} can be similarly applied to other high-resolution spectrographs, a well-validated library of spectra is required to be taken with the same instrument for maximum compatibility. \cite{HPFSM} adapted \texttt{HPFSpecMatch} for the NIR Habitable-zone Planet Finder \citep[HPF;][]{HPF_2} following the same methodology. \texttt{HPFSpecMatch} utilizes a library of HPF spectra comprising a subset of 166 stars from the \texttt{SpecMatch-Emp} library with $2770\,\text{K} < T_{\text{eff}} < 5990\,\text{K}$. 

We present \NEIDSM \citep[]{han_2025_14991481}, a spectra matching algorithm for the NEID spectrograph \citep{NEID_1}. The core of the code is migrated from \texttt{HPFSpecMatch} due to the shared pipeline architecture and data formats between NEID and HPF. We also reconstructed a library with NEID spectra for better compatibility and higher resolving power compared to the original HIRES spectra. 


\section{\NEIDSM Library} \label{sec:library}
\subsection{Library}
\begin{figure*}
    \centering
    \includegraphics[width=\textwidth]{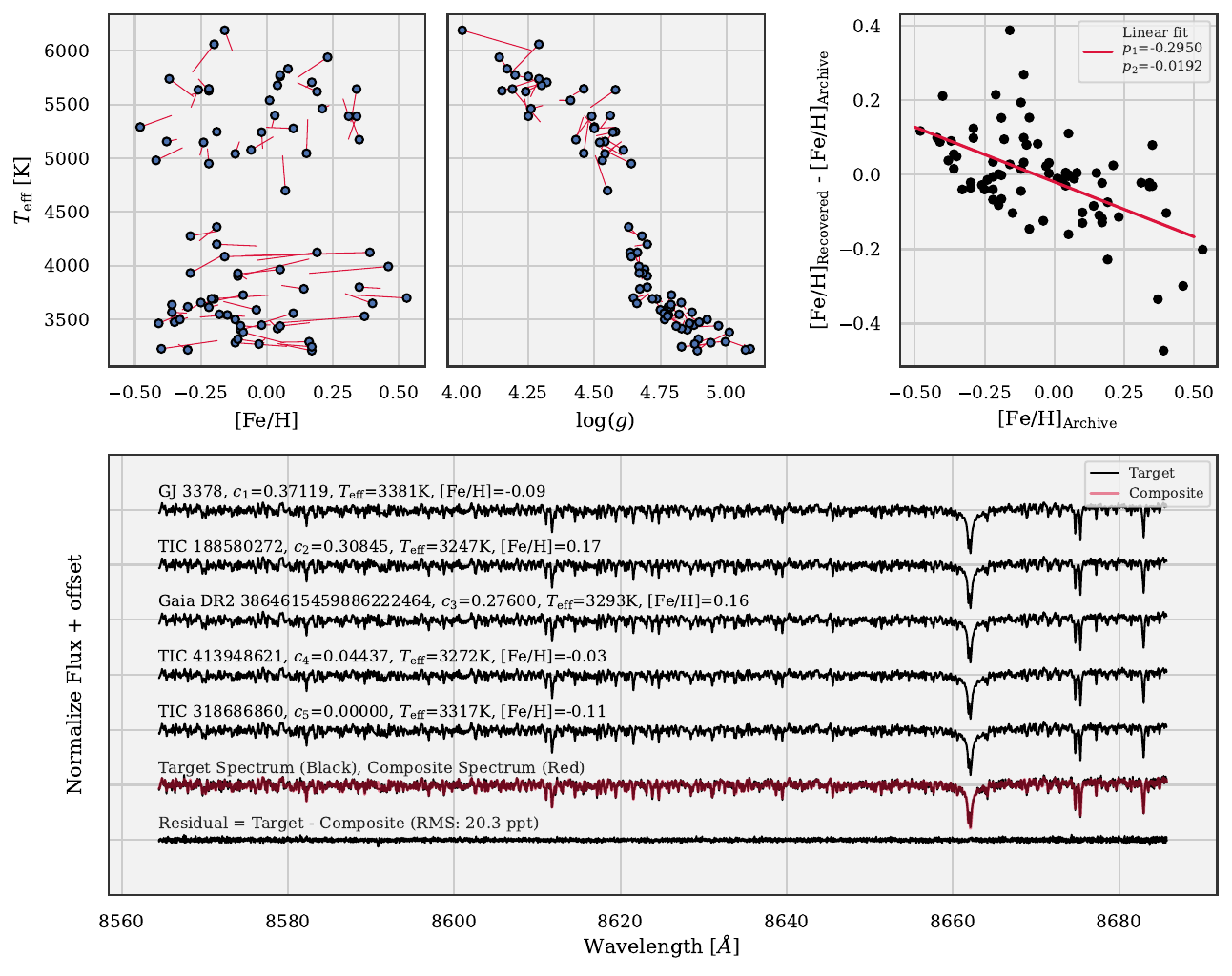}
    \caption{The \NEIDSM library distribution and TIC 437229644 spectral fit (order index 102). Top left: 78 \NEIDSM library stars in $T_{\mathrm{eff}}$, $\mathrm{[Fe/H]}$, and $\log g$ space: archival parameters (blue) vs. cross-validation results (red line endpoints). Top right: Residuals (recovered [Fe/H] - archival [Fe/H]) vs. archival [Fe/H].  Bottom: The composite spectrum (red)  derived with 5 best-fit library star spectra (black, weighted by $c_i$) plotted against the spectrum of TIC 437229644 (Target Spectrum). }
    \label{fig}
\end{figure*}
We searched the NEID Archive for the 404 stars from the library of the \texttt{SpecMatch-Emp} algorithm. By August 2024, 78 stars had been observed in the high resolution mode ($R \sim 120$k) with signal-to-noise ratio (S/N) greater than 50 at order index 102 ($\sim 8564 \text{\AA} - 8685 \text{\AA}$). All library spectra are reduced with NEID Data Reduction Pipeline v1.3.0, which applied an adaptive telluric model for order index 55 ($\sim 5174\text{\AA} - 5217 \text{\AA}$) to 103 ($\sim 8685\text{\AA} - 8810 \text{\AA}$). We only use the free spectral range\footnote{\url{https://neid.ipac.caltech.edu/docs/NEID-DRP/masterfiles.html}} of each order for spectra comparison. We selected the highest S/N 1-D extracted NEID spectrum of each star and adopted the stellar parameters determined by \cite{Yee}. The median uncertainties on the library's stellar parameters are $\bar\sigma_{T_{\text{eff}}}=60\,\text{K}$, $\bar\sigma_{\text{[Fe/H]}}=0.08$, and $\bar\sigma_{\log{g}}=0.05$. The first two panels of Figure \ref{fig} show the parameter space that the 78 library stars span. There is a gap at 4500 K due to a lack of observations, so we also analyze the library in two subsets: hot ($T_{\text{eff}} > 4500$ K) and cool ($T_{\text{eff}} < 4500$ K) stars. For hot stars, we have
\begin{equation}
    \bar\sigma_{T_{\text{eff}}}=56\,\text{K}, \bar\sigma_{\text{[Fe/H]}}=0.10, \text{and } \bar\sigma_{\log{g}}=0.03,
\end{equation}
and for the cool stars,
\begin{equation}
    \bar\sigma_{T_{\text{eff}}}=60\,\text{K}, \bar\sigma_{\text{[Fe/H]}}=0.08, \text{and } \bar\sigma_{\log{g}}=0.05.
\end{equation}
\subsection{Cross-validation}
We performed a cross-validation test following the method of \texttt{HPFSpecMatch} to examine the robustness and precision of the library. We removed each library star from the set, and fit its spectrum with the rest of the library stars in each NEID order between order index 55 and 103. The differences between the fitted and archival parameters---represented by the red lines in the first two panels of Figure \ref{fig}---indicate the precision of the current library. We calculated the standard deviation of the differences between the cross-validation-recovered parameters and the archival values as a measure of the uncertainty of the stellar parameter estimates derived from any spectrum on a certain order. Following cross-validation, we determined median parameter uncertainties across all orders to be
\begin{equation} \label{eq:median}
\sigma_{T_{\mathrm{eff}}} = 115\,\mathrm{K}, \sigma_{[\mathrm{Fe/H}]} = 0.143, \text{and } \sigma_{\log g} = 0.073. 
\end{equation}
The Mg I b triplet is known to be temperature sensitive for hot stars \citep{Yee}, which is included in order index 55, where we measure
\begin{equation} \label{eq:solar_err}
    \sigma_{T_{\mathrm{eff}}} = 110\,\mathrm{K}, \sigma_{[\mathrm{Fe/H}]} = 0.087, \text{and } \sigma_{\log g} = 0.091.
\end{equation} 
Redder NEID orders are generally measured with higher S/N for cool stars, and we showcase the uncertainties for order 102 (including the reddest line of the infrared Ca II triplet):
\begin{equation} \label{eq:M_err}
    \sigma_{T_{\mathrm{eff}}} = 64\,\mathrm{K}, \sigma_{[\mathrm{Fe/H}]} = 0.161, \text{and } \sigma_{\log g} = 0.042.
\end{equation}
In particular, the metallicity recovery for the cool stars exhibits higher uncertainty. To demonstrate the predictive power of the library stars on metallicity, we conducted a test in which we replaced the archival metallicity values with randomly selected values drawn from a distribution of the same size. The resulting recovered metallicity distribution showed larger scatter, with $\sigma_{[\mathrm{Fe/H}]} =0.26$. This increase in uncertainty confirms that the library metallicity values indeed possess predictive power. We note that there is a negative correlation between $\text{[Fe/H]}_{\text{archive}}$ and $\text{[Fe/H]}_{\text{recovered}} - \text{[Fe/H]}_{\text{archive}}$ according to the top right panel of Figure \ref{fig}. This is an expected artifact of the cross-validation method of the sparse non-uniform library: stars at the boundary of the library parameter spaces are biased towards the average values. 

\section{Application on M-dwarf Spectrum} \label{sec:app}
We demonstrate the application of \NEIDSM on an M-dwarf, highlighting the effectiveness on this challenging stellar population. Using TIC 437229644 as an example, we showcase the method's capability to estimate stellar parameters. At the bottom of Figure \ref{fig}, we show spectra of the five library stars with minimal differences (the smallest $\chi^2$ statistics) compared to the target spectrum. A composite spectrum from these five stars is shown atop the target spectrum, leaving a featureless residual. From order index 102, we measure 
\begin{equation}
        T_{\text{eff}} = 3311\,\text{K}, \text{[Fe/H]} = 0.088, \text{and } \log{g}=4.945,
\end{equation}
with uncertainties from Equation \ref{eq:M_err}. We also concluded an upper limit on $v \sin i$ of 2.8 km s$^{-1}$ for this target. This upper limit is set by the Line Spread Function (LSF) width of the instrument resolution for the high resolution mode. For $v \sin i$ fit results smaller than 2.8 km s$^{-1}$, it is recommended to quote this upper limit instead. For the high efficiency mode ($R \sim 70$k), the upper limit set by the resolution is $v \sin i < 4.5$ km s$^{-1}$. We also validated these limits by a $v \sin i$ injection-recovery test using the spectra of the slow-rotating Barnard's star, broadened by the methods of \cite{Carvalho2023}. 

For each order, \NEIDSM compiles result files including the $\chi^2$ values of each library star and result figures similar to the bottom panel of Figure \ref{fig}. Depending on the type and activity level of the star, the optimal spectral order with the highest S/N and fewest activity-sensitive lines varies. For example, a highly active M dwarf may not have ideal results using order index 102 because it includes one of the infrared Ca II triplet lines. The user must calculate the uncertainties of the chosen order by cross-validating or quote the inferred uncertainties of the newest library. The \NEIDSM \texttt{Github} repository\footnote{\url{https://github.com/TeHanHunter/neidspecmatch}} includes tutorials and library download links. \NEIDSM is also \texttt{pip}-installable\footnote{\url{https://pypi.org/project/neidspecmatch/}}.

\begin{acknowledgements}
    This work is supported by grant number 80NSSC23K0263 under the NASA Exoplanet Research Program.  
\end{acknowledgements}

\bibliography{neidsm}{}
\bibliographystyle{aasjournal}
\end{document}